\documentclass[12pt]{article}
\usepackage{latexsym,epsfig,epic,eepic}
\newcommand{\beq}{\begin{equation}}
\newcommand{\eeq}{\end{equation}}
\newcommand{\beqs}{\begin{eqnarray}}
\newcommand{\eeqs}{\end{eqnarray}}
\begin{document}
\begin{titlepage}
\vskip 1cm
\begin{center}
{\LARGE \bf Massless fermionic bound states}\\
\bigskip
{\LARGE \bf and the gauge/gravity correspondence}\\
\vspace{1.8cm} {\large Riccardo Argurio$^1$, Gabriele Ferretti$^2$ and\\
\smallskip
Christoffer Petersson$^2$} \vskip 0.7cm
{\large \it $^1$ Physique Th\'eorique et Math\'ematique\\
and\\
International Solvay Institutes\\
\smallskip
Universit\'e Libre de Bruxelles, C.P. 231, 1050 Bruxelles, Belgium\\
\vspace{0.5cm}
$^2$ Institute of Fundamental Physics \\
\smallskip
Chalmers University of Technology,\\
\smallskip
412 96 G\"oteborg, Sweden}

\end{center}
\vspace{1.5cm}

\begin{abstract}
We study the equations of motion of fermions in type IIB supergravity in
the context of the gauge/gravity correspondence.
The main motivation
is the search for normalizable fermionic zero modes in such backgrounds,
to be interpreted as composite massless
fermions in the dual theory.
We specialize to
backgrounds characterized by a constant dilaton and a self-dual
three-form. In the specific case of the Klebanov--Strassler
solution we construct explicitly the fermionic superpartner of
the Goldstone mode associated with the broken baryonic symmetry.
The fermionic equations could also
be used to search for goldstinos in theories that break
supersymmetry dynamically.
\end{abstract}

\end{titlepage}

\section{Introduction and summary}
One of the latest surges of interest in the context of the
gauge/gravity correspondence (for reviews close to the topics of this
work, see
\cite{Aharony:2002up,Bertolini:2003iv,Strassler:2005qs})
has been the possibility that
some backgrounds might provide
the supergravity realization of dynamical supersymmetry (SUSY) breaking.
This possibility was first considered for the quiver theories described
in~\cite{Berenstein:2005xa,Franco:2005zu,Bertolini:2005di}.
These theories were constructed as a non-conformal
deformation of the conformal
theories~\cite{Martelli:2004wu,Bertolini:2004xf, Benvenuti:2004dy}
dual to the new Sasaki-Einstein
manifolds~\cite{Gauntlett:2004yd,Gauntlett:2004hh} $Y^{pq}$.
Unfortunately, in spite of being chiral theories, these
theories do not display true dynamical supersymmetry breaking with a
stable ground state, but rather a runaway
behavior~\cite{Franco:2005zu,Intriligator:2005aw} very much
like super QCD with $0<N_f<N_c$~\cite{Affleck:1983mk}. Still, the
possibility of the existence of
gravity solutions dual to dynamically broken SUSY has not been
ruled out. (Some work on deformations
for these theories can be found
in~\cite{Herzog:2004tr,Burrington:2005zd,Pal:2005mr,Sfetsos:2005kd,Berg:2005pd}.
For related earlier work see~\cite{Page:1985bq}.)

One of the consequences of spontaneous SUSY breaking
(dynamical~\cite{Witten:1981nf} or tree
level~\cite{Fayet:1974jb, O'Raifeartaigh:1975pr}) is the existence of a
fermionic Goldstone
mode $g$ -- the  ``goldstino''~\cite{Salam:1974zb}.
Such mode can arise as a massless bound state of microscopic degrees
of freedom in a
confining theory and has the distinguishing property of coupling to the
supercurrent $J$ without derivative
terms -- in obvious notation:
\beq
        \langle 0|J^\mu_\alpha|g_\beta \rangle = f \gamma^\mu_{\alpha\beta},
\label{theproperty}
\eeq
where $f\not=0$ is the goldstino coupling.
In the context of the gauge/gravity correspondence, such particle must
be described by a normalizable zero
mode in the bulk coupling directly to the gravitino $\Psi_\mu$ and can
be studied by looking at the bulk
fermionic equations of motion.

Even in theories that do not break SUSY, the study of the bulk fermionic
equations and the
search for normalizable zero modes is still of interest. Obviously, with
unbroken SUSY, the bosonic
and fermionic spectra must match and one does not obtain additional
information from the latter. However,
particularly in the case of zero modes, some information may be easier
to obtain in the second case, since
the fermionic equations are easily linearized and index theorems may be
available. In some cases, such
as the cascading theory of Klebanov and Strassler
(KS)~\cite{Klebanov:2000hb} massless fermionic
modes (the ``axino'')\footnote{Strictly speaking, one should refrain to call
such multiplet ``axionic'' since it is not related to an anomalous symmetry,
like the QCD axion. Still, we will, in a few places, use the word
within quotes for sake of brevity and to make connection with the
previous literature.}
must exist as a superpartner to the Goldstone
boson associated with the breaking of the baryonic $U(1)$
symmetry~\cite{Aharony:2000pp,Gubser:2004qj,Gubser:2004tf} and their
explicit construction strengthens the correspondence.
More generically, ${\cal N}=1$ SUSY implies the presence of massless
fermionic
superpartners of the scalar fields parameterizing the quantum moduli space
of vacua, when there is one
(some of these scalars can be seen as Goldstone bosons of broken
global continuous symmetries).

The ``axino'' does not obey
(\ref{theproperty}) (since SUSY is unbroken in this case) and its explicit
form helps elucidating
precisely how (\ref{theproperty}) should be interpreted in the bulk.
The solution that we find in section~4 has the property that it does
not give a source for the supercovariant field strength of the gravitino,
more specifically
\beq
      \Gamma^{MNP}\mathcal{D}_N \Psi_P = 0 ~~~~~ \hbox{on
shell}.\label{thecon}
\eeq
(The notation is discussed in section~2). We propose that the signature of
spontaneous SUSY breaking is the existence of a normalizable zero mode
for which (\ref{thecon}) is not satisfied.

Another way to distinguish a generic massless fermion from the goldstino
is by looking at how they transform under the global symmetries of the
problem\footnote{We thank I.~Klebanov for pointing out this possibility
to us in the context of the KS solution.}. For instance, the bosonic
zero modes found in~\cite{Gubser:2004qj} are odd under
the $Z_2$ symmetry exchanging the two $S^2$ spheres of the deformed
conifold and the same symmetry should act non-trivially on their fermionic
superpartner. On the other hand, a true goldstino should be invariant
under such symmetries. We will discuss the details for the KS solution
in the conclusions after having presented the explicit solution.

The purpose of this paper is twofold -- on the one hand, we wish
to begin addressing the general issues above for a class of KS-like
backgrounds
(consisting of a constant dilaton and a self-dual three-form) and, on the
other, we test these techniques in the
true KS model~\cite{Klebanov:2000hb} and
construct explicitly the fermionic zero mode.
Eventually, one will have to consider more complicated backgrounds with
more general fluxes
but we feel that the class we are considering in this paper is a good
starting point
to sharpen one's tools and includes at least the important example
of~\cite{Klebanov:2000hb}.
Various aspects of flux compactifications that might be relevant in this
context are reviewed
in~\cite{Grana:2005jc}.

Perhaps the most interesting quality of the general
equations we discuss is that the existence of a zero mode
hinges on the existence of a solution to the massless Dirac equation on
a (six-dimensional)
Ricci flat manifold (see section~3.4).
In the compact case, the existence of such a
solution implies the existence of a
covariantly constant spinor and thus of a K\"ahler structure, by the
standard arguments
of integration by part. In the non-compact case however, the boundary
terms cannot be neglected and,
because of the presence of the warp factor, there is a possibility for
having a normalizable zero
mode without necessarily implying a K\"ahler structure. We shall discuss
this possibility in the conclusions, after having presented the
dependence of the equations from the warp factor.
Another possibility would be to leave the K\"ahler structure
untouched but change the three-form appropriately.

The paper is organized as follows: In section~2 we begin by reviewing
the fermionic equations of
motion of type IIB supergravity obtained in~\cite{Schwarz:1983qr}
(see
also~\cite{Schwarz:1983wa,Howe:1983sr,Kavalov:1986ki,Castellani:1993ye}).
In section~3
we specialize to the above mentioned class of backgrounds and show how
the equations for the zero modes
can be reduced to a set of Dirac and Rarita-Schwinger equations on the
internal manifold, starting
precisely with the Dirac equation discussed above.
In section~4 we
turn to an application of the equations just derived
and use them to construct the
``axino'' for the true KS solution. This zero mode is the fermionic partner
of the Goldstone mode associated
to the breaking of the baryonic $U(1)$ symmetry and is not to be thought
as a goldstino and in fact condition (\ref{thecon}) is satisfied.
We briefly summarize our findings in section~5 and present a more detailed
discussion of the $Z_2$ symmetry transformations of the KS solution and
comment on the issue related to the K\"ahler structure mentioned above.
Some useful formulas, like the explicit expression for the spin connection
on the deformed conifold, are collected in the appendix.

\section{The fermionic equations of motion of type IIB supergravity}

In this section we review the fermionic equations of motion of type IIB
supergravity
obtained in~\cite{Schwarz:1983qr}. This
allows us to make some comments on the conventions and notation used. We
will set the Newton constant
to one, $\kappa=1$, for convenience. (It can always be reinstated by
dimensional analysis.)
We will only work to first order in the fermionic fields.

In order to follow the
more recent literature, we will use, contrary to~\cite{Schwarz:1983qr}, a
``mostly plus'' metric.
This can be most easily accomplished by letting $g_{MN} \to -g_{MN}$,
$\Gamma_M \to i \Gamma_M$ and
so on, and implies a few sign changes that are easily implemented. The
$\Gamma$-matrices are all
real in the Majorana representation.

Our convention for the $\epsilon$-tensor is that it includes the
appropriate
determinant of the metric and thus
transforms as a true tensor, not as a density. Also, when evaluated with
flat
indices it is purely numerical and we have the sign convention
$\epsilon_{{0}\dots {9}} = - \epsilon^{{0}\dots {9}} = 1$.
Finally, the five-form $F_5$ is self-dual ($*_{10} F_5 = F_5$) in the sense
\beq
      F_{M_1 M_2 M_3 M_4 M_5} = \frac{1}{5!}\epsilon_{M_1 M_2 M_3 M_4
M_5 M_6 M_7 M_8 M_9 M_{10}}
      F^{M_6 M_7 M_8 M_9 M_{10}}
\eeq

We also define, with flat indices, $ \Gamma_{\chi 10} = \Gamma^{{0}}
\dots \Gamma^{{9}}$,
and the chiralities of the dilatino $\lambda$ and gravitino $\Psi_M$ are:
$\Gamma_{\chi 10} \lambda = - \lambda$,  $\Gamma_{\chi 10} \Psi_M = +
\Psi_M$,
reversed from the conventions in~\cite{Schwarz:1983qr}.

The dilatino and gravitino equations of motion are,
respectively~\cite{Schwarz:1983qr}:
\beqs
     \Gamma^M D_M \lambda &=& \frac{i}{240}\Gamma^{MNPQR}F_{MNPQR}
\lambda +
     \frac{1}{24}\Gamma^M \Gamma^{NPQ}G_{NPQ} \Psi_M  \nonumber \\
     &&+ \Gamma^M \Gamma^R P_R \Psi_M^*, \label{eqlambda}
\eeqs
and
\beqs
     &&\Gamma^{MNP} D_N \Psi_P = -\frac{1}{48} \Gamma^{NRL} \Gamma^M
     G^*_{NRL} \lambda -
     \frac{i}{480} \Gamma^{MNP} \Gamma^{QRLST} F_{QRLST} \Gamma_N \Psi_P
\nonumber \\
     &&\;\;\;\;\;\;\;\ + \frac{1}{96} \Gamma^{MNP}
        (\Gamma_N^{\phantom{N}LSR} G_{LSR} - 9 \Gamma^{LS} G_{NLS})
       \Psi_P^* + \frac{1}{2}\Gamma^R \Gamma^M P_R \lambda^*. \label{eqpsi}
\eeqs

Note that the $\Psi$ terms on the RHS of (\ref{eqlambda}) and
(\ref{eqpsi}) are not written out
explicitly in~\cite{Schwarz:1983qr} but they are certainly present for
supercovariance as can be seen by taking
the SUSY variations, which in our notation read:
\beq
    \delta \lambda= \Gamma^M P_M \varepsilon^* + \frac{1}{24}
\Gamma^{MNP}G_{MNP}  \varepsilon,
    \label{susydil}
\eeq
and
\beq
    \delta \Psi_M = D_M  \varepsilon + \frac{i}{480} \Gamma^{NPQRS}F_{NPQRS}
    \Gamma_M  \varepsilon - \frac{1}{96} ({\Gamma_M}^{NPQ}G_{NPQ} -9
\Gamma^{NP}
    G_{MNP})  {\varepsilon^*}.  \label{susygrav}
\eeq
We have chosen to write out explicitly all the fermionic terms to avoid
confusion but one could just as well introduce a supercovariant
derivative $\mathcal{D}_N$ in terms of which eq. (\ref{eqpsi}) becomes
simply
\beq
    \Gamma^{MNP} \mathcal{D}_N \Psi_P = -\frac{1}{48} \Gamma^{NRL} \Gamma^M
     G^*_{NRL} \lambda + \frac{1}{2}\Gamma^R \Gamma^M P_R \lambda^*.
    \label{supercov}
\eeq
The RHS of (\ref{supercov}) acts as a source for the supercovariant field
strength of the gravitino.

The ordinary covariant derivatives are by definition:
\beqs
      D_M \lambda &=&
      \left(\partial_M + \frac{1}{4} \omega_M^{AB}
      \Gamma_{AB} - \frac{3}{2} i Q_M\right)\lambda\\
      D_M \Psi_R &=& \left(\partial_M +
      \frac{1}{4} \omega_M^{AB} \Gamma_{AB} - \frac{1}{2} i Q_M
\right)\Psi_R -
      \mathbf{\Gamma}_{MR}^L \Psi_L,
\eeqs
where $Q_M$ is the auxiliary $U(1)$ field introduced
in~\cite{Schwarz:1983qr} and $\omega_M^{AB}$ the
usual spin connection.
Notice that the contribution of the Christoffel symbol $
\mathbf{\Gamma}_{MR}^L$ drops out
in the kinetic term for the gravitino but, without it, the derivative is
no longer covariant.

\section{The fermionic equations of motion in a KS-like ansatz}

We now specialize the equations reviewed in the previous section to a
generic KS-like background
precisely defined as follows.

\subsection{Bosonic ansatz}
Let us review the ansatz step by step in order to distinguish between
the basic assumptions and their consequences. We start from the 4+6 split
of the geometry. The ten dimensional metric is split into a
four-dimensional
warped Minkowski space described by the
coordinates $x^\mu$ and a six-dimensional internal space described by
the coordinates $y^i$:
\beq
  ds_{10}^2 = e^{-\frac{1}{2}u(y)} dx^\mu  dx_\mu +
e^{\frac{1}{2}u(y)}  d\hat{s}^2
\eeq
where $e^{u(y)}$ is the warp factor and $d\hat{s}^2 = g_{ij}(y)
dy^i dy^j$ is the internal
metric which is assumed to describe a
smooth non-compact manifold.

To avoid confusion we stress that the
six-dimensional indices $i,j\dots$ will always be raised/lowered with
the metric
$g_{ij}$ and all powers of the warp factor written explicitly. Also,
with a slight
abuse of notation, the covariant derivative $D_i$ will denote the true
covariant
derivative on the internal manifold, and thus it is shifted (by a term
containing the warp factor) with respect to the
one used in section~2. A subtlety that arises when commuting it through
a $\Gamma$-matrix is discussed in appendix~A.

By Poincar\'e invariance in the Minkowski space, all other fields can
depend only on the $y^i$ coordinates. Furthermore, the complex 3-form $G_3$
must be living purely in the six-dimensional internal space.

The basic assumption that we make is to take the
3-form to be imaginary self-dual in the six-dimensional internal space:
\beq
     *_6 G_3 = i G_3,
\label{isd}
\eeq
that is
\beq
     \frac{1}{6}\epsilon_{ijklmn}
     G^{lmn} = i G_{ijk},
\eeq
where the $\epsilon$-tensor is defined with respect to the internal metric
$g_{ij}$.
In particular, for flat indices we have
$\epsilon_{{4}\dots {9}} = \epsilon^{{4}\dots {9}} = 1 $.

The assumption (\ref{isd}) leads to many simplifications.
First of all, we can consider a background where the type IIB dilaton and
RR scalar can be held constant,
thus allowing us to set
\beq
        P_M = Q_M = 0.
\eeq
We can think of this condition as a kind of extremality condition, since
the equations of motion for the dilaton and axion are sourceless for our
ansatz.
The Bianchi identities further impose that the 3-form is closed:
\beq
dG_3=0,
\eeq
and self-duality thus requires it to be harmonic.

To preserve 4d Poincar\'e symmetry the self-dual 5-form must be taken as:
\beq
F_5 = {\cal F}_5 + *_{10} {\cal F}_5, \qquad \qquad {\cal F}_5= {\cal
F}_1 \wedge
dx^0\wedge dx^1 \wedge dx^2 \wedge dx^3.
\eeq
The equations of motion of the 5-form are:
\beq
dF_5 = \frac{1}{8} i G_3\wedge G^*_3. \label{eomF5}
\eeq
Since $G_3$ is purely in the 6-manifold, the EOM above imply that
$d{\cal F}_5 = 0$ and thus ${\cal F}_1 = dZ$, with $Z=Z(y)$ a real function.

Now the EOM for the 3-form are:
\beq
d*_{10}G_3= 4 iF_5\wedge G_3.
\eeq
Taking into account self-duality of $G_3$, we have:
\beq
*_{10}G_3 = i e^{-u} G_3 \wedge dx^0\wedge dx^1 \wedge dx^2 \wedge
dx^3.
\eeq
Thus the EOM for $G_3$ imply that $d(4 Z-e^{-u})\wedge G_3=0$ over
the 6-manifold, which, due to self-duality of $G_3$ implies that
$Z=\frac{1}{4}e^{-u}$ up to an additive constant that we set to zero.
Thus:
\beq
{\cal F}_5= \frac{1}{4}d e^{-u} \wedge
dx^0\wedge dx^1 \wedge dx^2 \wedge dx^3.
\eeq
Note that the sign of the 5-form is directly related to the sign
in the self-duality equation for $G_3$.

The Einstein equations for the metric, given the above source fields,
yield for the internal part just the condition that the 6-dimensional
metric is Ricci-flat, $R_{ij}=0$. For the 4-dimensional part, they yield
an equation for the warp factor that can be also consistently derived
from the EOM of the 5-form (\ref{eomF5}) with indices along the 6-manifold,
and which is entirely determined by the data on the
six-dimensional manifold:
\beq
      - \nabla_6 e^u =
      \frac{1}{12} G^*_{lmn}
      G^{lmn}  \label{telodo}
\eeq

Of course, the above equations do not imply
SUSY. As well known~\cite{Grana:2000jj}, SUSY requires in addition
the internal space to be K\"ahler and the three-form
to be $(2,1)$ and primitive. The KS background obeys these conditions
and is thus supersymmetric.
However, we will not make this assumption in our derivation, except in
section~4 where we shall specialize to the KS background.

Without (or even with) SUSY, one might wonder whether it makes sense to
impose
the self-duality condition on the 3-form. Relaxing this condition
would imply a much more generic, but also much more complicated, set up.
Though such a generalization should ultimately be carried out, we feel
that the above set up is first of all a good training ground, but might
also be
of relevance in situations in which SUSY is present
asymptotically, and the 3-form could well preserve its
self-duality everywhere.

\subsection{Fermionic ansatz}
Now it is time to introduce a 4-6 split for the spinors and the
$\Gamma$-matrices.
We split the $\Gamma$-matrices as follows
\beq
    \Gamma^\mu = e^{\frac{u}{4}} \gamma^\mu\otimes \mathbf{1}, \quad
    \Gamma^i = e^{-\frac{u}{4}}\gamma_{\chi 4}\otimes \gamma^i,
    \quad\mathrm{with}\quad
    \gamma_{\chi 4} = i \gamma^0\dots\gamma^3.
\eeq
The warp factors have been denoted explicitly so that the four and
six dimensional $\gamma$-matrices
obey
\beq
     \{ \gamma^\mu,  \gamma^\nu \} = 2 \eta^{\mu\nu} \quad\mathrm{and}\quad
     \{ \gamma^i,  \gamma^j \} = 2 g^{ij}.
\eeq
We are in a Majorana representation where all $\gamma^\mu$ are real and
all $\gamma^i$ imaginary.
Similar equations, with the warp factors reversed,
hold for $\Gamma_\mu$ and $\Gamma_i$.
We also define, with flat indices,
\beq
    \gamma_{\chi 6} = -i \gamma^{{4}} \dots \gamma^{{9}},
\eeq
which is such that $\Gamma_{\chi 10} = \gamma_{\chi 4}\otimes
\gamma_{\chi 6}$ .

We consider one of the two
linearly independent constant Weyl spinors in four dimensions $\epsilon_+$
of positive four dimensional
chirality  together with its complex conjugate $\epsilon_- = \epsilon_+^*$
of negative four dimensional
chirality.
We make the most general ansatz that is suited to search
for zero momentum massless modes with four-dimensional spin $1/2$:
\beqs
    \lambda &=& \epsilon_+ \otimes \lambda_- + \epsilon_- \otimes \lambda_+
\nonumber \\
    \Psi_\mu &=& \Gamma_\mu(\epsilon_+ \otimes \chi_- + \epsilon_- \otimes
\chi_+) \nonumber \\
    \Psi_i &=&  e^{\frac{u}{4}}(\epsilon_+ \otimes \psi_{+i} +
\epsilon_-
\otimes \psi_{-i}). \label{splitfe}
\eeqs
The $\pm$ signs denote the four and six dimensional chiralities
and in the case of $\lambda$
we use the same symbol for the six dimensional spinor as for the
ten dimensional one since no confusion can
arise. The warp factor in the last of (\ref{splitfe}) has been
introduced for convenience.
Notice that, apart from  $\epsilon_\pm$, the other spinors are
not the complex conjugate of each other since the ten dimensional
fermions are not Majorana.

\subsection{Fermionic equations of motion, preliminaries}
It is now straightforward to insert (\ref{splitfe}) and the bosonic ansatz
into the equations of motion (\ref{eqlambda}), (\ref{eqpsi})
and to collect the terms proportional to $\epsilon_+$ and
those proportional to $\epsilon_-$.
We obtain equations that contain only data from the six dimensional
manifold. Namely, the dilatino equation (\ref{eqlambda})
gives rise to the following two equations:
\beq
    \gamma^i{D}_i \lambda_- + \frac{3}{8}
    \gamma^i \partial_i u \lambda_- =
    \frac{1}{4} e^{-\frac{u}{2}} \gamma_{jk} {G}^{ijk} \psi_{+i}
\label{eomlm}\eeq
and:
\beq
    \gamma^i{D}_i \lambda_+ - \frac{1}{8}
    \gamma^i \partial_i u \lambda_+ =
    -\frac{1}{6} e^{-\frac{u}{2}} \gamma_{ijk} {G}^{ijk} \chi_+.
\label{eomlp}\eeq
Similarly, the component along $x^\mu$ of (\ref{eqpsi}) gives rise to:
\beqs
    &&\gamma^{ij}{D}_i \psi_{+j}
    - \frac{1}{2} {\partial}^i u \psi_{+i}
    + \frac{3}{8} \gamma^{ij} \partial_i u \psi_{+j}
    + 3 \gamma^i {D}_i \chi_-
    - \frac{3}{8} \gamma^i \partial_i u \chi_-
    \nonumber \\
    &&= \frac{1}{48} e^{-\frac{u}{2}} \gamma^{ijk}
    G^*_{ijk} \lambda_- + \frac{1}{8} e^{-\frac{u}{2}}
    {G}^{nij}\gamma_{ij}\psi^*_{-n}
\label{eomchim}\eeqs
and:
\beqs
    &&\gamma^{ij}{D}_i \psi_{-j}
    + \frac{3}{8} \gamma^{ij} \partial_i u \psi_{-j}
    - 3 \gamma^i {D}_i \chi_+
    - \frac{9}{8} \gamma^i \partial_i u \chi_+
   \nonumber \\
    &&= \frac{1}{8} e^{-\frac{u}{2}} \gamma^{ijk}
    G_{ijk} \chi^*_- - \frac{1}{8} e^{-\frac{u}{2}}
    {G}^{nij}\gamma_{ij}\psi^*_{+n}
\label{eomchip}\eeqs

Finally, the component along $y^i$ of (\ref{eqpsi}) yields:
\beq
    \gamma^{pij}{D}_i \psi_{+j} - \frac{1}{8}\gamma^{pij}
    \partial_i u \psi_{+j} +
    4 \gamma^{pi} {D}_i \chi_- -\frac{1}{2} \gamma^{pi}
    \partial_i u \chi_-
    = \frac{1}{2} e^{-\frac{u}{2}} {G}^{pij} \gamma_{ij} \chi^*_+
\label{eompsip}\eeq
and
\beqs
    &&\gamma^{pij}{D}_i \psi_{-j} + \frac{3}{8}\gamma^{pij}
    \partial_i u \psi_{-j} -
    4 \gamma^{pi} {D}_i \chi_+ + \frac{1}{2} \gamma^{pi}
    \partial_i u \chi_+ - 2 {\partial}^p u \chi_+ \nonumber \\
    &&= \frac{1}{8} e^{-\frac{u}{2}} {G}^{*pij} \gamma_{ij} \lambda_+ +
    \frac{1}{2} e^{-\frac{u}{2}} {G}^{pij} \gamma_{ij} \chi^*_- -
    \frac{1}{2} e^{-\frac{u}{2}} {G}^{pij} \gamma_i \psi^*_{+j}
\label{eompsim}\eeqs

Before making any further manipulation, it is advisable to check which of
the six dimensional fermions can or cannot
be gauged away in this particular bosonic background.
Therefore we reserve to the SUSY variations
(\ref{susydil}) and  (\ref{susygrav}) the same treatment
we gave the equations of motion. For the
SUSY variation:
\beq
    \varepsilon = \epsilon_+ \otimes \varepsilon_+ +
    \epsilon_- \otimes \varepsilon_-:
\eeq
(where, again, $\varepsilon_+$ and  $\varepsilon_-$ are independent), we
get:
\beqs
     \delta \lambda_+ &=& 0 \nonumber \\
     \delta \lambda_- &=& \frac{1}{24} e^{-\frac{3u}{4}}
     G_{ijk} \gamma^{ijk} \varepsilon_+\nonumber \\
     \delta \chi_+ &=& 0 \nonumber \\
     \delta \chi_- &=& - \frac{1}{4} e^{-\frac{u}{4}}
     \gamma^i \partial_i u \varepsilon_+
          - \frac{1}{96} e^{-\frac{3u}{4}} G_{ijk}
          \gamma^{ijk} \varepsilon^*_- \\
     e^{\frac{u}{4}} \delta \psi_{+i} &=& {D}_i \varepsilon_+
           + \frac{1}{4} \gamma_{ij} {\partial}^j u\varepsilon_+
           - \frac{1}{8} \partial_i u \varepsilon_+ + \frac{1}{16}
           e^{-\frac{u}{2}} G_{ijk} \gamma^{jk} \varepsilon_-^*
\nonumber \\
     e^{\frac{u}{4}} \delta \psi_{-i} &=&
     {D}_i \varepsilon_- + \frac{1}{8} \partial_i u \varepsilon_-
           + \frac{1}{8} e^{-\frac{u}{2}} G_{ijk}
           \gamma^{jk} \varepsilon_+^* \nonumber
\eeqs

The usual gauge choice
$\Gamma^M \psi_M=0$ can be easily seen to correspond to
$4 \chi_- + \gamma^i \psi_{+i} =0$ and $4 \chi_+ - \gamma^i \psi_{-i} =0$,
but we will choose a more convenient one in the following.

\subsection{Disentangling the fermionic equations of motion}
We now rewrite all the fermionic equations as a system which can be
solved step by step, in principle by inverting the Dirac operator on the
6 dimensional transverse manifold.

First of all, we subtract from (\ref{eomchip}) the contraction with
$\gamma_p$
of (\ref{eompsim}), to obtain the massless Dirac equation discussed
in the introduction:
\beq
\gamma^i {D}_i \tilde \chi_+ =0, \qquad \mathrm{where}\qquad
\tilde\chi_+ = e^{-\frac{5}{8}u}
\chi_+ . \label{starthere}
\eeq
In order to rewrite (\ref{eompsip}), we define
\beq
\tilde \psi_{+i} =  e^{-{\frac{u}{8}}}( \psi_{+i} +\gamma_i \chi_-).
\eeq
If we then choose the gauge $\gamma^i \tilde \psi_{+i} =0$,
we obtain the following simple equation:
\beq
\gamma^j {D}_j \tilde \psi_{+i} = \frac{1}{2} G_{ijk}
\gamma^{jk} \tilde\chi_+^*.\label{notsosimple}
\eeq
Note that contracting with $\gamma^i$ we obtain the condition
(that was used to obtain the previous equation):
\beq
{D}^i  \tilde \psi_{+i} = 0.
\eeq
Now we turn to (\ref{eomlm})-(\ref{eomlp}). We perform the rescalings:
\beq
\tilde\lambda_+ = e^{-\frac{u}{8}}  \lambda_+ , \qquad \qquad
\tilde\lambda_- = e^{\frac{3}{8}u}\lambda_-.
\eeq
Then the equations simply write:
\beqs
\gamma^i {D}_i \tilde \lambda_+ &=& -\frac{1}{6} G_{ijk}
\gamma^{ijk} \tilde \chi_+ ,\label{delpiu} \\
\gamma^i {D}_i \tilde \lambda_- &=& \frac{1}{4} {G}^{ijk}
\gamma_{jk} \tilde \psi_{+i}. \label{delmeno}
\eeqs
Turning to (\ref{eompsim}), we define:
\beq
\tilde\psi_{-i} = e^{\frac{3}{8}u} \psi_{-i} .
\eeq
Then, imposing the gauge $\gamma^i \tilde \psi_{-i} =0$,
we obtain the equation:
\beq
\gamma^j {D}_j \tilde \psi_{-i} = - 4 e^u {D}_i \tilde\chi_+
-  \gamma_{ij} {\partial}^j e^u \tilde\chi_+ -
\partial_i e^u \tilde\chi_+
+\frac{1}{8} G^*_{ijk} \gamma^{jk} \tilde \lambda_+
-\frac{1}{2} G_{ijk} \gamma^j \tilde \psi_{+}^{*k}.  \label{almostthere}
\eeq
For the sake of completeness, the contraction with $\gamma^i$ gives:
\beq
{D}^i \tilde \psi_{-i} = -3 \gamma^i \partial_i e^u
\tilde\chi_+.
\eeq
We are left with (\ref{eomchim}). After we perform an additional rescaling:
\beq
\tilde \chi_-= e^{\frac{7}{8}u}  \chi_-,
\eeq
the equation becomes:
\beq
\gamma^i {D}_i  \tilde \chi_- = -\frac{1}{2} \partial^i
e^u \tilde \psi_{+i}
-\frac{1}{96} G^*_{ijk} \gamma^{ijk} \tilde \lambda_-
-\frac{1}{16} {G}^{ijk} \gamma_{ij} \tilde \psi_{-k}^*.
\label{chimenotilde}
\eeq
Note that the SUSY variation of the gauge fixing conditions is simply
given by:
\beqs
\gamma^i \delta \tilde \psi_{+i} & = & \gamma^i {D}_i \tilde \varepsilon_+,
\qquad \qquad \varepsilon_+ = e^{\frac{3}{8}u} \tilde \varepsilon_+, \\
\gamma^i \delta \tilde \psi_{-i} & = & \gamma^i {D}_i \tilde \varepsilon_-,
\qquad \qquad \varepsilon_- = e^{-\frac{u}{8}} \tilde \varepsilon_-.
\eeqs

\section{Finding an explicit fermionic solution in the KS background}
In this section, we apply the equations derived above to study the
problem of finding a fermionic massless zero mode in the
supersymmetric KS background~\cite{Klebanov:2000hb}. The existence
of such mode is needed in order to form a SUSY multiplet together
with the two bosonic massless modes (sometimes referred to as the
``axion'' and the ``saxion'' with a slight abuse of language) which
have been derived
in~\cite{Gubser:2004qj,Gubser:2004tf}\footnote{After the bosonic
modes were given, the full baryonic branch was constructed
in~\cite{Butti:2004pk}, using the techniques of~\cite{Grana:2004bg}
and showed to agree with the gauge theory analysis
in~\cite{Dymarsky:2005xt}. Another deformation, which breaks SUSY
explicitly, was considered
in~\cite{Kuperstein:2003yt,Schvellinger:2004am}.}. The ``axion'' is
actually the Goldstone boson associated with the breaking of the
baryonic symmetry~\cite{Aharony:2000pp,Gubser:2004qj,Gubser:2004tf}.
Finding the ``axino'' completes the holographic description of the
massless multiplet present in the low energy effective description
of the boundary theory, and is thus a nice check of the
gauge/gravity correspondence.

We first review the KS background~\cite{Klebanov:2000hb}.
This is just a specific case
of the generic ansatz discussed in section~3.1 and thus,
we only need to know that the internal space is the deformed conifold
\cite{Candelas:1989js} (see
also~\cite{Minasian:1999tt,Ohta:1999we,Arean:2004mm}),
whose sechsbein are, up to an overall rescaling:
\beqs
    e^1 &=& A(\tau)(-\sin\theta_1 d\phi_1 - \cos\psi\sin\theta_2 d\phi_2
    + \sin\psi d\theta_2 ) \nonumber \\
    e^2 &=& A(\tau)(d\theta_1 - \sin\psi \sin\theta_2 d\phi_2 - \cos\psi
    d\theta_2) \nonumber \\
    e^3 &=& B(\tau)(-\sin\theta_1 d\phi_1 + \cos\psi\sin\theta_2
    d\phi_2 - \sin\psi d\theta_2 ) \nonumber \\
    e^4 &=& B(\tau)(d\theta_1 + \sin\psi \sin\theta_2 d\phi_2 +
    \cos\psi d\theta_2) \label{sechsbein} \\
    e^5 &=& C(\tau)(d\psi + \cos\theta_1 d\phi_1 + \cos\theta_2
    d\phi_2) \nonumber \\
    e^6 &=& C(\tau) d\tau \nonumber
\eeqs
where, in terms of the function:
\beq
    K(\tau) = \frac{(\sinh \tau \cosh \tau - \tau)^{1/3}}{\sinh\tau},
\eeq
defined in~\cite{Klebanov:2000hb}, we have:
\beqs
     A^2(\tau) &=& \frac{1}{4}K(\tau) (\cosh \tau -1), \nonumber \\
     B^2(\tau) &=& \frac{1}{4}K(\tau) (\cosh \tau +1), \\
     C^2(\tau) &=& \frac{1}{3 K^2(\tau)}. \nonumber
\eeqs
For the sake of completeness, we give the spin connection in appendix~A.
We can then write the equations for a covariantly constant spinor on
the deformed conifold, $D_i \eta=0$. With our coordinate choice
(\ref{sechsbein}), they imply that the spinor must be constant, and has
to obey the conditions (with flat indices):
\beq
     (\gamma^{12} + \gamma^{34})\eta =
     (\gamma^{16} - \gamma^{45})\eta = 0 \label{covcov}
\eeq
We will choose $\eta$ to have positive
chirality $\gamma_{\chi6}\eta = \eta$ and denote its complex conjugate
(of negative chirality) by $\eta^*$.
Then, $\eta$ satisfies three conditions:
\beq
(\gamma^1 +i\gamma^4)\eta=0, \qquad (\gamma^3 +i\gamma^2)\eta=0, \qquad
(\gamma^6 -i\gamma^5)\eta=0. \label{projreal}
\eeq
The above formula allows one to read off the complex structure
in the flat indices. Denoting complex indices in boldface, we take the
following holomorphic sechsbein:
\beq
e^\mathbf{1} = e^1+ie^4, \qquad e^\mathbf{2} = e^3+ie^2, \qquad
e^\mathbf{3} = e^6-ie^5, \label{complexstructure}
\eeq
and the antiholomorphic sechsbein are obtained by complex conjugation.
The unconventional combinations above are forced upon us
by the labeling of the sechsbein
(\ref{sechsbein}) that is the one commonly used in the literature.
With these normalizations, the flat metric is
$\eta^{\mathbf{1}\bar\mathbf{1}} = 2$ and
$\eta_{\mathbf{1}\bar\mathbf{1}} = 1/2$.

With (\ref{complexstructure}), the conditions (\ref{projreal}) simply
become:
\beq
     \gamma^\mathbf{a} \eta = 0, \qquad  \mathbf{a}=\mathbf{1},
     \mathbf{2},\mathbf{3}.
\eeq

Using the covariantly constant spinor $\eta$, we can now start
solving the fermionic equations in the KS background.

Eq. (\ref{starthere}) can be trivially solved by setting:
\beq
\tilde \chi_+ = \eta. \label{andthus}
\eeq
Notice that the
expression for  $\chi_+ = e^{\frac{5}{8}u} \eta$
is then normalizable in the sense of~\cite{Henningson:1998cd}
(see also~\cite{Muck:1998iz,Arutyunov:1998ve,Henneaux:1998ch}, the case
for the Rarita-Schwinger field is discussed
in~\cite{Corley:1998qg,Volovich:1998tj,Rashkov:1999ji,Matlock:1999fy}).

We already notice from the asymptotic behavior of the solution above,
that the mode we have found should be dual to an operator of dimension
$\Delta=\frac{5}{2}$, which is the right dimension for the fermion
in the ``axion-saxion'' chiral multiplet, which has dimension
$\Delta=2$, see~\cite{Gubser:2004qj}.

Moving on to (\ref{notsosimple}), we need, first of all,
an expression for the three form $G_3$. This
is given in~\cite{Klebanov:2000hb} and has the form
(in the flat basis (\ref{complexstructure})):
\beqs
G_3&= &  \sqrt{3} M\left[\frac{(\tau \cosh\tau-\sinh\tau)}{\sinh^3\tau}
e^\mathbf{1}\wedge e^\mathbf{2} \wedge e^\mathbf{\bar 3} \right.
\label{gi} \\ && \left.
+ \frac{(\sinh \tau \cosh\tau -\tau)}{2\sinh^3\tau}
(e^\mathbf{\bar 1}\wedge e^\mathbf{2} \wedge e^\mathbf{3}
-e^\mathbf{1}\wedge e^\mathbf{\bar 2} \wedge e^\mathbf{3} )
\right], \nonumber
\eeqs
where $M$ is the number of fractional branes in the KS set up.
In relation to the complex structure
(\ref{complexstructure}) $G_3$ is indeed a
$(2,1)$ primitive form, so that $G_{ijk} \gamma^{ijk} \eta=0$.

It can also be easily checked that the RHS of (\ref{notsosimple})
has only antiholomorphic indices and
thus it is consistent\footnote{
Indeed, the spin connection is such that the covariant derivative
does not couple holomorphic and antiholomorphic indices
(with respect to the flat basis
(\ref{complexstructure})).}
to take, in the flat basis:
\beq
\tilde\psi_{+\mathbf{a}}=0.
\eeq
Making now the ansatz that $\tilde\psi_{+i}$
depends explicitly only on $\tau$ one can show (by requiring the
$\theta_1$ dependence of (\ref{notsosimple}) to cancel out algebraically)
that the most general form for the remaining components is:
\beqs
      \tilde\psi_{+\mathbf{\bar 1}} &=& z \gamma^\mathbf{1} \eta^*
      +   v \gamma^\mathbf{2} \eta^* \nonumber \\
      \tilde\psi_{+\mathbf{\bar 2}} &=& v \gamma^\mathbf{1} \eta^*
      +   z \gamma^\mathbf{2} \eta^* \label{ansapiu} \\
      \tilde\psi_{+\mathbf{\bar 3}} &=&   -2 z \gamma^\mathbf{3} \eta^*
      \nonumber
\eeqs
The terms proportional to $z(\tau)$ are solutions of an homogeneous
equation whereas
$v(\tau)$ couples to the source.
The remaining conditions are all solved by the functions
\beqs
     z(\tau) &=& \frac{c}{\sinh\tau\cosh\tau  - \tau} \\
     v(\tau) &=& -M \frac{(\tau\cosh\tau-\sinh \tau)}
     {K \sinh^2\tau}.\nonumber
\eeqs
Requiring  $\tilde\psi_{+i}$ to be regular at the origin sets
$c = 0$.\footnote{
We can actually write the solution above
in closed form, as $\tilde\psi_{+i} = -2i B_{ij} \gamma^j \eta^*$,
where $B_2$ is the 2-form potential of the imaginary part of $G_3$,
also given in \cite{Klebanov:2000hb}.
Note that $B_2$ is a $(1,1)$ primitive form
which satisfies $d*_6 B_2 =0$, conditions which are necessary for
consistency with (\ref{notsosimple}).}
Normalizability can be checked using the boundary
terms discussed
in~\cite{Corley:1998qg,Volovich:1998tj,Rashkov:1999ji,Matlock:1999fy}.

Armed with the explicit solutions of (\ref{starthere}) and
(\ref{notsosimple})
we can easily solve the dilatino
equations (\ref{delpiu}) and (\ref{delmeno}).
The source for (\ref{delpiu}) is identically zero and
normalizability forces us to take
\beq
       \lambda_+ = 0
\eeq
The source for (\ref{delmeno}) turns out to be proportional to
$\gamma^\mathbf{3}\eta^*$ times an overall
function of $\tau$ allowing for the ansatz
$\tilde\lambda_- = f(\tau)\eta^*$.
Inserting in (\ref{delmeno}) we find $ f= 4 e^u$,
implying the normalizability of
\beq
       \lambda_- = 4 e^{\frac{5}{8}u}\eta^* = 4 \chi_+^*.
       \label{fassino}
\eeq
Notice that any dependence on $\lambda_-$ disappears in the
remaining equations.

It is now time to look at (\ref{almostthere}).
It will be useful to have the explicit expression for the warp factor.
With the normalizations
(\ref{sechsbein}) and (\ref{gi}), we have, from (\ref{telodo}):
\beq
     e^{u(\tau)}  = 2 M^2 \int^\infty_\tau d\tau^\prime
     K(\tau')\frac{(\tau^\prime\coth\tau^\prime -1)}
     {\sinh \tau^\prime}. \label{effi}
\eeq

In this case the source term has both holomorphic
and antiholomorphic indices and the resulting set of equations cannot
be solved in terms of
elementary functions. Still, it is possible to completely characterize
the solution and its asymptotic behavior in terms of the warp factor.
Making the ansatz that $\tilde\psi_{-i}$ only depends
explicitly on $\tau$ and imposing the gauge condition,
one can write $\tilde\psi_{-i}$ in terms of
three unknown functions:
\beqs
     && \tilde\psi_{-\mathbf{1}} = r(\tau) \gamma^\mathbf{\bar 1} \eta,\quad
        \tilde\psi_{-\mathbf{2}} = r(\tau) \gamma^\mathbf{\bar 2} \eta,\quad
        \tilde\psi_{-\mathbf{3}} =
                 -2 r(\tau) \gamma^\mathbf{\bar 3} \eta,\nonumber \\
     && \tilde\psi_{-\mathbf{\bar 1}} = s(\tau) \gamma^\mathbf{\bar 1}
        \eta,\quad
        \tilde\psi_{-\mathbf{\bar 2}} = - s(\tau)
        \gamma^\mathbf{\bar 2} \eta,\quad
        \tilde\psi_{-\mathbf{\bar 3}} = t(\tau) \gamma^\mathbf{\bar 3} \eta.
           \label{vabbe}
\eeqs
One could also add to (\ref{vabbe}) a solution of the homogeneous equation,
similar to the $z(\tau)$
dependence of (\ref{ansapiu}) that decouples from the system and
should be set to zero anyway by
imposing regularity at the origin and normalizability.
Inserting (\ref{vabbe}) into (\ref{almostthere}) yields three linear
first order O.D.E.s in the
three unknown functions $r,s,t$ that can be further simplified
into two decoupled O.D.E.s
(one of first order and the other of second order) for $r$ and $s$:
\beq
    r^\prime(\tau) + \frac{2 \sinh^2 \tau}{\cosh\tau \sinh\tau - \tau}
    r(\tau) =  \frac{1}{2} \partial_\tau e^{u(\tau)},
\eeq
and:
\beq
     s^{\prime\prime}(\tau) + 4 \coth\tau  s^{\prime}(\tau) + 3 s(\tau) =
       - \frac{1}{2} \frac{\partial_\tau e^{u(\tau)}}{\sinh\tau},
\eeq
where we have used (\ref{effi}).
Furthermore, $t$ is expressed in terms of $s$:
\beq
t(\tau)= (s(\tau) \sinh \tau)'.
\eeq
The reason why the equations for $s(\tau)$ and
$r(\tau)$ decouple is that one can solve separately the equations for the
holomorphic and antiholomorphic components.

After some manipulations it turns out that they can both be solved
in terms of simple integrals, much like the
warp factor:
\beqs
      r(\tau) &=& \frac{1}{2}
      e^{u(\tau)} - \frac{1}{\sinh\tau\cosh\tau - \tau}
       \int_0^\tau d\tau^\prime  e^{u(\tau^\prime)}
       \sinh^2\tau^\prime \nonumber \\
        &=& -M^2 \frac{1}{\sinh\tau\cosh\tau - \tau}
        \int_0^\tau d\tau^\prime K^4(\tau') \sinh \tau'
        (\tau' \cosh\tau' -\sinh\tau'), \nonumber \\
      s(\tau) &=& -\frac{1}{2\sinh^3\tau} \int_0^\tau d\tau^\prime
      e^{u(\tau^\prime)}
        \sinh^2\tau^\prime, \label{rs} \\
        &=& \frac{1}{4} K^3 (\tau) (2r(\tau) - e^{u(\tau)}), \nonumber
\eeqs
where all the integration constants have been fixed so that
the solution is regular at the origin and
normalizable. Note that $s(\tau)$ is asymptotically subleading, so that
the components of the solution depending explicitly on it vanish faster
as the
boundary is approached.

At last, we conclude this derivation by finding
the expression for $\chi_-$ from
(\ref{chimenotilde}). Inserting the expression for $\tilde\psi_{-k}^*$
in the RHS we find that the source is
pointing along $\gamma^\mathbf{1}\gamma^\mathbf{2}\gamma^\mathbf{3}\eta^*$.
Perhaps the most convenient way to write the source is in terms
of the function $s(\tau)$ in (\ref{rs}) and its derivative:
\beq
    \gamma^i D_i  \tilde \chi_- =
    \frac{\sqrt{3}M}{4 \sinh\tau}(\tau s(\tau) +
           (\tau\coth\tau -1)s'(\tau))
           \gamma^\mathbf{1}\gamma^\mathbf{2}\gamma^\mathbf{3}\eta^*.
\label{finita}
\eeq
(Notice that
$ \gamma^\mathbf{1}\gamma^\mathbf{2}\gamma^\mathbf{3}\eta^* \propto \eta $.)
This suggests the ansatz
\beq
    \tilde \chi_- = w(\tau) \gamma^\mathbf{1}
    \gamma^\mathbf{2}\eta^*  \label{prodi}
\eeq
 and in fact, (\ref{finita}) yields:
\beq
       w(\tau) = \frac{M}{2}\frac{\tau\coth\tau -1}{K(\tau)\sinh\tau}
s(\tau),
\eeq
where, once again, the integration constant has been fixed by the
requirement that the solution be regular at
the origin. One can check that $\chi_-$ is also normalizable.

This completes the finding of the zero momentum massless fermionic mode.
Having found the explicit solution it is very easy to check that
condition (\ref{thecon}) is obeyed (in the sense that the RHS of
(\ref{supercov}) vanishes) due to the simple expression for the dilatino.

\section{Conclusions}
In this paper we began a systematic study of the fermionic equations of
motion
of IIB supergravity in the context of the gauge/gravity correspondence
with emphasis to the search for bulk zero modes dual to massless fermions
in the gauge theory. We stressed that among all such fermions, the one
associated with SUSY breaking (if it occurs) should be singled out by
looking at the gravitino fluctuation, most likely through the
contribution to its supercovariant field strength.
Other fermionic massless modes, such as the
KS ``axino'' do not contribute to this quantity.
It is interesting to note that the vanishing of the RHS of
(\ref{supercov}), at least for our ansatz, is closely related to the
conditions for SUSY preservation, namely $G_{ijk} \gamma^{ijk} \eta=0$.
This is no longer vanishing even
in the mildest way to break SUSY, i.e.
by the presence of a $(0,3)$ piece for $G_3$.

Another way to distinguish a generic massless fermion from the goldstino
is by looking at how it transforms under the global symmetries of the
problem. For instance, the bosonic
zero modes found in~\cite{Gubser:2004qj,Gubser:2004tf} are odd under
the $Z_2$ symmetry exchanging the two $S^2$ spheres of the deformed
conifold and the same symmetry should act non-trivially on its fermionic
superpartner.

Let us briefly recall the origin of this symmetry.
The exchange of the two
spheres is implemented by the exchange of the pairs of coordinate
$(\theta_1, \phi_1)$ and $(\theta_2, \phi_2)$ in the solution.
Trivially, the sechsbein $e^5$ and $e^6$ in (\ref{sechsbein})
are invariant under the exchange but the remaining four transform in a
complicated way. One can construct however various combinations that
transform in a simple way:
\beq
    (e^1)^2 +  (e^2)^2 \quad\hbox{and}\quad
    (e^3)^2+  (e^4)^2
\eeq
that are even under the exchange, and
\beq
    e^1 \wedge e^2, \quad  e^3 \wedge e^4\quad\hbox{and}\quad
     e^1 \wedge e^3 +  e^2 \wedge e^4  \label{bertinotti}
\eeq
that are odd.
Hence, of the bosonic fields, the (constant) dilaton, the metric and the
five-form are even while the three form $G_3$ is odd.
Let us also recall that the bosonic
zero mode $a(x)$ constructed in~\cite{Gubser:2004qj} enters as
$\delta G_3 = *_4 da + \dots$ and thus it must be odd under the symmetry
so as to preserve the overall parity of $G_3$.

Let us now look at the fermionic solution presented in section~4.
The transformation properties of the fields $\psi_{+i}$ and $\psi_{-i}$
are somewhat complicated by the fact that they carry an
internal index but we don't really need them for
the argument -- it is quite enough to look at
$\chi_\pm$ and $\lambda_\pm$.

The claim is that $\chi_+$ and $\lambda_-$ are even while $\chi_-$ is
odd ($\lambda_+$ is zero).
To see this, notice that the covariantly constant spinors $\eta$
and $\eta^*$ are both even because the six-dimensional chirality
is unchanged by the symmetry (one is exchanging two pairs of indices).
Thus $\chi_+$ and $\lambda_-$ given in (\ref{andthus}) and
(\ref{fassino}) are even\footnote{The quantities with a tilde only
differ by powers of the warp factor and have thus the same transformation
properties.}.
On the other hand, expanding the solution (\ref{prodi}) for $\chi_-$
in terms of the gamma matrices in the real basis, one gets a spinor
proportional to
\beq
   \left(\gamma^1 \gamma^3 + \gamma^2 \gamma^4 +
   i (\gamma^1 \gamma^2 - \gamma^3 \gamma^4)\right) \eta^*
\eeq
that transforms as the combinations of sechsbein in (\ref{bertinotti})
and it is thus odd.
To compensate for that in the expression for $\Psi_\mu$
we must let the zero mode
$\epsilon_\pm \to   - \gamma_{\chi 4} \epsilon_\pm$,
thus showing that it transforms non-trivially under the $Z_2$ symmetry.

The second point briefly mentioned in the
introduction that we would like to discuss is the possibility
that the presence of the warp factor might allow for normalizable
zero modes without requiring a K\"ahler structure and thus SUSY.

There are two different types of
boundary conditions that should be considered. Let us begin with the
standard one. Assume that one is looking at an
internal manifold whose metric is asymptotically that of a cone:
\beq
       d\hat{s}^2 \approx dr^2 + r^2 d \Sigma^2,
\eeq
for some five-dimensional Einstein (but not necessarily Sasaki)
manifold with metric $d\Sigma^2$. The existence of a covariantly
constant spinor would of course imply a K\"ahler structure for the
manifold, but the original condition (the Dirac equation (\ref{starthere}))
necessary for the zero mode is weaker on a non-compact
manifold. The two conditions are equivalent by the standard
argument of integration by part only for spinors $\tilde\chi_+$
vanishing at infinity faster that $r^{-2}$. (The covariantly constant
spinor $\eta$ is an exception because it is completely independent on $r$.)

Would the existence of a spinor solving the
Dirac equation (\ref{starthere}) but decaying more slowly than $r^{-2}$
still allow for a massless mode on the boundary? For this we must look
at the other boundary condition inferred from the AdS/CFT correspondence
in~\cite{Henningson:1998cd}. Namely, we must ensure that, for instance:
\beq
    \left.\left( \sqrt{G}\bar{\chi}_+ \chi_+
    \right)\right|_{\hbox{bdry}} < \infty, \label{goran}
\eeq
where $G$ is the determinant of the induced metric at the boundary.
This is one of the conditions that has been used throughout section~4 to
check
for normalizability.
Inserting the appropriate powers of the warp factor $e^u \approx r^{-4}$,
one sees that (\ref{goran}) requires $\tilde\chi_+$ to scale like
$r^\alpha$ with $\alpha < 1/2$. Thus, the possibility of having
a normalizable zero mode without a K\"ahler structure is
left open.

Whether gravity duals to theories with a stable non-supersymmetric vacuum
exist is still an open and interesting question. It seems that
one would need the background to be dual
to a chiral gauge theory with no classical flat directions
(see the arguments and caveats in \cite{Affleck:1984xz}). The cascading
theories considered until now are not of this kind, and we do not expect a
smooth gravity dual for a theory with no stable vacuum. Presumably,
one will have to turn to a more general ansatz, but we hope that a similar
analysis to the one performed in this paper can be helpful in this
endeavor.

\subsection*{Acknowledgments}

It is a pleasure to thank M.~Bertolini, F.~Bigazzi and A.~L.~Cotrone for
many discussions at the beginning of this project. In particular,
M.~Bertolini
has been extremely helpful in elucidating many aspects of the gauge/gravity
correspondence to us while this work was being carried out.
We are very grateful to I.~Klebanov for reading a draft of
the manuscript and making suggestions on how to improve it.
We also benefited from conversations with S.~Kuperstein, D.~Martelli,
L.~Martucci and A.~Zaffaroni
and from email exchanges with I.~Pesando and D.~Tsimpis.

R.A. is a Research Associate of the Fonds National de la Recherche
Scientifique (Belgium).
The research of R.A. is partially supported by IISN - Belgium (convention
4.4505.86), by the ``Interuniversity Attraction Poles Programme --
Belgian Science Policy'' and by the European Commission FP6
programme MRTN-CT-2004-005104, in which he is associated to
V.U.Brussel.
The research of G.F. is supported by the Swedish Research Council
(Vetenskapsr{\aa}det) contracts 622-2003-1124 and 621-2002-3884.
Partial support from the EU Superstring Theory Network, project number
MRTN-CT-2004-512194 is also gratefully acknowledged.

\section*{Appendix {\bf A}}

We collect some useful formulas that have been used extensively in the
derivation of the equations in the text.

Let us begin with the spin connection.
In the flat basis, the non-zero components of the spin
connection $\omega^{ab,c} = - \omega^{ba,c}$ are
\beqs
    && \omega^{12,1} = \omega^{34,1} = \frac{\cos\theta_1}
    {2 A(\tau) \sin\theta_1} \nonumber \\
    && \omega^{16,1} = -\omega^{45,1} = \omega^{26,2} = \omega^{35,2}
    = f_1(\tau) \nonumber \\
    && \omega^{12,3} = \omega^{34,3} = \frac{\cos\theta_1}{2 B(\tau)
      \sin\theta_1} \nonumber \\
    && \omega^{36,3} = -\omega^{25,3} = \omega^{46,4} = \omega^{15,4} =
    f_2(\tau)  \label{spinconnection}\\
    && \omega^{12,5} = \omega^{34,5} = \frac{1}{2 C(\tau)} \nonumber \\
    && \omega^{23,5} = -\omega^{14,5} = \frac{1}{2} \omega^{56,5} = f_3
    (\tau) \nonumber
\eeqs
where in defining the functions $f_1, f_2, f_3$ we have taken
into account the following identities:
\beqs
       f_1(\tau) &=& \frac{A^\prime(\tau)}{A(\tau) C(\tau)} =
                     \frac{-A^2(\tau)+B^2(\tau)+C^2(\tau)}
                     {4 A(\tau)B(\tau)C(\tau)}\nonumber \\
       f_2(\tau) &=& \frac{B^\prime(\tau)}{B(\tau) C(\tau)} =
                     \frac{A^2(\tau)-B^2(\tau)+C^2(\tau)}
                     {4 A(\tau)B(\tau)C(\tau)}\\
       f_3(\tau) &=& \frac{C^\prime(\tau)}{2 C^2(\tau)} =
                     \frac{A^2(\tau)+ B^2(\tau)-C^2(\tau)}
                     {4 A(\tau)B(\tau)C(\tau)}.\nonumber
\eeqs

Other useful formulas are those involving the self-dual forms $G_3$ and
$F_5$:
\beqs
      \gamma^{ijk} G_{ijk} &=&  \gamma^{ijk} G_{ijk}\frac{1 +
       \gamma_{\chi 6}}{2} \nonumber \\
      \gamma^m\gamma^{ijk} G_{ijk} &=&  6 \gamma_{ij} G^{mij}\frac{1 +
         \gamma_{\chi 6}}{2} \nonumber \\
      \gamma^{ijk} \gamma^m G_{ijk} &=&  6 \gamma_{ij} G^{mij}\frac{1 -
          \gamma_{\chi 6}}{2},
\eeqs
and similarly for the complex conjugates (recall that $\gamma_i$ and
$\gamma_{\chi 6}$ are all imaginary
in order for the ten dimensional matrices to be real).
The terms containing the five-form on the other hand, can be simplified
with the help
of:
\beq
    \frac{i}{240} \Gamma^{MNPQR}F_{MNPQR}  =
    \frac{1}{4}\partial_mu \Gamma^m \Gamma_{\chi 6} \frac{1 -
\Gamma_{\chi 10}}{2},
\eeq
where $\Gamma_{\chi 10} = \Gamma^{{0}} \dots \Gamma^{{9}}$ as in the
text and
$\Gamma_{\chi 6} = -i \Gamma^{{4}} \dots \Gamma^{{9}} = \mathbf{1}\otimes
\gamma_{\chi 6}$ all with
flat indices. (See also e.g.~\cite{Kehagias:1998gn,Gubser:2000vg}.)

Lastly, recall that, since $D_i$ represents the covariant
derivative on the internal manifold,
it no longer commutes with $\Gamma_\mu$ and we have instead:
\beq
     D_i \Gamma_\mu =  \Gamma_\mu D_i - \frac{1}{4} \partial_i u
\Gamma_\mu.
\eeq

\end{document}